\documentclass[prl,aps,twocolumn,nobibnotes,nofootinbib]{revtex4}
\usepackage{graphicx}
\usepackage{amsmath}
\usepackage{amssymb}
\usepackage{bm}
\usepackage{color}

\begin{document}

\title{Non-Markovian spin relaxation in two-dimensional electron gas. }
\author{M.M.~Glazov}
\address{A. F. Ioffe Physico-Technical Institute, Russian Academy of Sciences,
194021 St. Petersburg, Russia}
\author {
E.Ya.~Sherman}
\address{Department of Physics and Institute for Optical
Sciences, University of Toronto, 60 St. George Street, Toronto,
Ontario, Canada M5S 1A7}

\begin{abstract}
We analyze by Monte-Carlo simulations and analytically spin
dynamics of two-dimensional electron gas (2DEG) interacting with
short-range scatterers in nonquantizing magnetic fields. It is
shown that the spin dynamics is non-Markovian with the exponential
spin relaxation followed by the oscillating tail due to the
electrons residing on the closed trajectories. The tail relaxes on
a long time scale due to an additional smooth random potential and
inelastic processes. The developed analytical theory and
Monte-Carlo simulations are in the quantitative agreement with
each other.

Journal ref.: Europhys. Lett. {\bf 76}, pp. 102-108 (2006).
\end{abstract}

\maketitle

Spin-dependent phenomena in semiconductors and semiconductor
nanostructures
attract an increasing interest during last decade (for review see, e.g. Ref.~%
\onlinecite{zutic:323}). The understanding of the mechanisms of
spin decoherence, usually considered as a Markovian process, as
well as the possibilities to control it is an important problem in
the field. In the two-dimensional (2D) zinc-blende structures the
electron spin dynamics is governed by the spin-orbit (SO)
splitting of the conduction band. It originates from the lack of
the inversion center either in the bulk material as represented by
the bulk inversion asymmetry (BIA or Dresselhaus
term~\cite{dresselhaus55,dyakonov86}) or the structural asymmetry
of the heteropotential revealed by the Rashba
term~\cite{rashba64}. In all the cases the SO interaction gives
rise to the effective Zeeman magnetic field which is characterized
by the electron wavevector $\bm k$-dependent spin precession
vector $\bm\Omega _{\bm k}$ whose direction determines the axis
and the absolute value is the spin precession rate. In the
collision dominated regime of electron motion with $\Omega \tau
\ll 1$
(where $\Omega $ is the typical value of $\left\vert \bm\Omega _{\bm %
k}\right\vert $ and $\tau $ is the scattering time) the small spin
rotation angles $\Omega \tau $ between successive collisions are
not correlated, thus the spin relaxation rate $\tau _{s}^{-1}$ is
proportional to the $\Omega ^{2}\tau
$.~\cite{dyakonov72,dyakonov86} This D'yakonov-Perel' spin
relaxation mechanism, which represents a Markovian process, is the
most important one in the wide range of temperatures and carrier
concentrations.~\cite{zutic:323,ivchenko05a,
averkiev02,malinowski00,karimov03,brand:236601,song.66.035207,ohno.83.4196}
At given sample parameters there are two possibilities to control
the spin relaxation rate with this mechanism. First, one may apply
an external electric field in order to tune the Rashba constant
~\cite{karimov03,miller03}. Second, an external magnetic field can
suppress the spin relaxation due to the cyclotron rotation, Larmor
spin precession or their
combination.~\cite%
{ivchenko73,wilamowski:035328} Although the main mechanisms of the
magnetic field effects on the Markovian spin relaxation are
understood, \cite{glazov04} the peculiarities in the spin dynamics
caused by the fascinating non Markovian transport phenomena in
magnetic field, lavished attention only very recently
\cite{pershin04,lyubinskiy04,glazov05}, and are scarcely known.
Non-Markovian spin dynamics can arise for electrons interacting
with nuclear spins \cite{merkulov02}, hovewer, with a
qualitatively different mechanism of the memory.

\begin{figure}[htbp]
\includegraphics[width=0.8\linewidth]{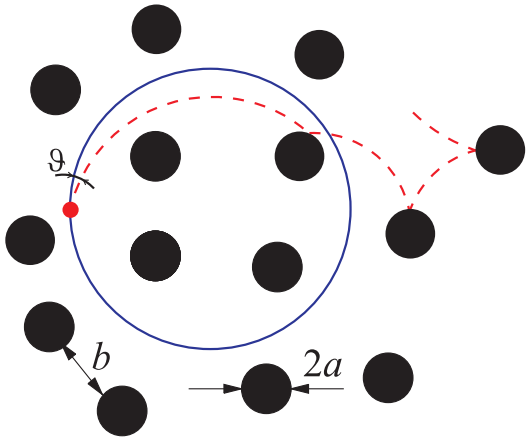}
\caption{Electron's motion along the circling trajectory. Black
circles -- scattering centers. Solid line -- initial ``circling''
trajectory. Dashed line --  wandering trajectory after the
inelastic scattering event with the scattering angle $\vartheta$.
Here $a$ is the scatterer radius and $b\sim N^{-1/2}$ is the mean
distance between scatterers. Figure shown not to scale.}
\label{fig:scatter}
\end{figure}

In this Letter we point out and study a new non-Markovian effect
in electron spin relaxation in the magnetic field, namely, the
appearance of the long-living spin polarization tail. We consider
a degenerated 2DEG with SO coupling scattered by the short-range
impurity centers (or antidots) with the radius $a$ and the
concentration $N$ in the presence of the magnetic field ${\bf B}$
directed perpendicular to the structure plane. Assuming specular
reflection by the scatterers, the kinetics of electrons is
characterized by  the total scattering length $l=1/(2Na)$ and the
transport mean free path $l_{tr}=3l/4$. As shown in the pioneering
work ~\cite{baskin78} (see also
Refs.~\onlinecite{baskin98,dmitriev02,bobylev97} and references
therein) a fraction of electrons $P_c (B)$ moving with the
velocity $v$ does not experience scattering travelling along the
closed circular orbits. In the $Na^2\ll 1$ limit it is given by
\begin{equation}
P_c (B)=\exp {(-2\pi R_{c}/l)},  \label{wp}
\end{equation}%
where $R_{c}=v/\omega_{c}$ is the cyclotron radius,
$\omega_{c}=eB/mc$ is the cyclotron frequency, $e$ is the
elementary charge and $m$ is the electron effective mass.
Physically, Eq. \eqref{wp} gives the probability to find no
scatterers in the area between circles of the radii $R_{c}\pm a$, see Fig.~%
\ref{fig:scatter}. As the electron finished the cyclotron
revolution without a collision it will stay on this orbit
infinitely long, thus demonstrating a non Markovian behavior with
an infinitely long memory. The fraction of these electrons
increases with the magnetic field resulting in the classical
magnetoresistance~\cite{dmitriev02}. Here we discuss the role of
the closed orbits in the spin relaxation. We show that the spin
dynamics of the electrons on such trajectories is strictly
periodical and the total spin falls exponentially down to the
value determined by the fraction of circular orbits. Therefore, a
tail in the spin polarization can be observed. Further, we develop
a theory describing a decay of this tail caused by the scattering
processes which transfer electrons between closed and open
``wandering"~\cite{dmitriev02} orbits. Finally, we discuss the
role of the quantum effects.

In order to obtain quantitative results we consider a
$\hat{z}\parallel [001]$-grown zince-blende quantum well (QW) and
assume that the SO interaction is dominated by the Dresselhaus
term~\cite{dyakonov86}
\begin{equation}
\mathcal{H}_{SO}(\bm k)=\frac{\hbar }{2}(\bm\sigma \cdot \bm\Omega
_{\bm k}), \qquad
\bm\Omega _{\bm k}=\Omega _{0}[\cos {\varphi _{\bm k}},-\sin {\varphi _{\bm %
k}},0]. \label{hso}
\end{equation}%
where $\bm\sigma =(\sigma _{x},\sigma _{y},\sigma _{z})$ is the
Pauli matrix vector, $\Omega _{0}$ is the spin precession
frequency at the Fermi level and $\varphi _{\bm k}$ is the angle
between electron wavevector $\bm k$ and $\hat{x}\parallel [100]$
axis.

We assume classical dynamics of the electrons and their spins and
subdivide the electrons into two groups. First one corresponds to
wandering electrons while the second one corresponds to circling
electrons, their relative fractions are $1-P_c (B)$ and $P_c (B)$,
respectively. The accurate description of the wandering electrons
in the moderate magnetic fields can be done by using the kinetic
equation and their spin decays exponentially with the
rate~\cite{ivchenko73}
\begin{equation}
\frac{1}{\tau _{s}(B)}=\frac{\Omega _{0}^{2}\tau }{1+(\omega
_{c}\tau )^{2}}, \label{taus}
\end{equation}%
where $\tau =l_{tr}/v_{F}$. In the derivation of Eq. \eqref{taus}
we assumed $\Omega _{0}\tau \ll 1$~ (\footnote{This condition can
be relaxed provided $\Omega _{0}\ll \omega _{c}$, M.M. Glazov, to
be published}) and neglected the Larmor effect on spin relaxation
as it is usually small in GaAs-based QWs. The Larmor effect can be
taken into account in a standard way.~\cite{glazov04}

The spin of the single electron with $\bm k(t)$-dependent
wavevector is described by equation
\begin{equation}
\frac{\partial \bm s}{\partial t}+\bm s\times \bm\Omega _{\bm
k(t)}=0. \label{dsdt}
\end{equation}%
Eq. \eqref{dsdt} can be readily solved assuming that  $\bm k$
rotates in the QW plane with the frequency $\omega _{c}$. For the
initial condition $2{\bm s}(0)=(0,0,1)$, we obtain for
$z$-component
\begin{equation}
2s_{z}(t)=\frac{\omega _{c}^{2}}{\omega _{c}^{2}+\Omega _{0}^{2}}+\frac{%
\Omega _{0}^{2}}{\omega _{c}^{2}+\Omega _{0}^{2}}\cos {\left(
\sqrt{\Omega _{0}^{2}+\omega _{c}^{2}}t\right) }.  \label{szt}
\end{equation}%
The solutions for $s_{x,y}$ are rather cumbersome and not
presented here. It is seen in Eq. \eqref{szt} that on the circling
trajectory $2s_z$ oscillates around the mean value
$\omega_{c}^{2}/(\omega _{c}^{2}+\Omega _{0}^{2})\rightarrow 1$
(in high fields, $%
\omega _{c}\gg \Omega _{0}$) with the frequency $\sqrt{\Omega
_{0}^{2}+\omega _{c}^{2}}$, and the amplitude of the oscillations
is $\Omega _{0}^{2}/(\omega _{c}^{2}+\Omega _{0}^{2})\rightarrow
0$ at $\omega _{c}\gg \Omega _{0}$. We note that these
oscillations are insensitive to the initial phase of the $\bm k$
rotation and thus all the spins of the circling electrons
oscillate synchronously.

\begin{figure*}[htbp]
\includegraphics[width=0.4\linewidth]{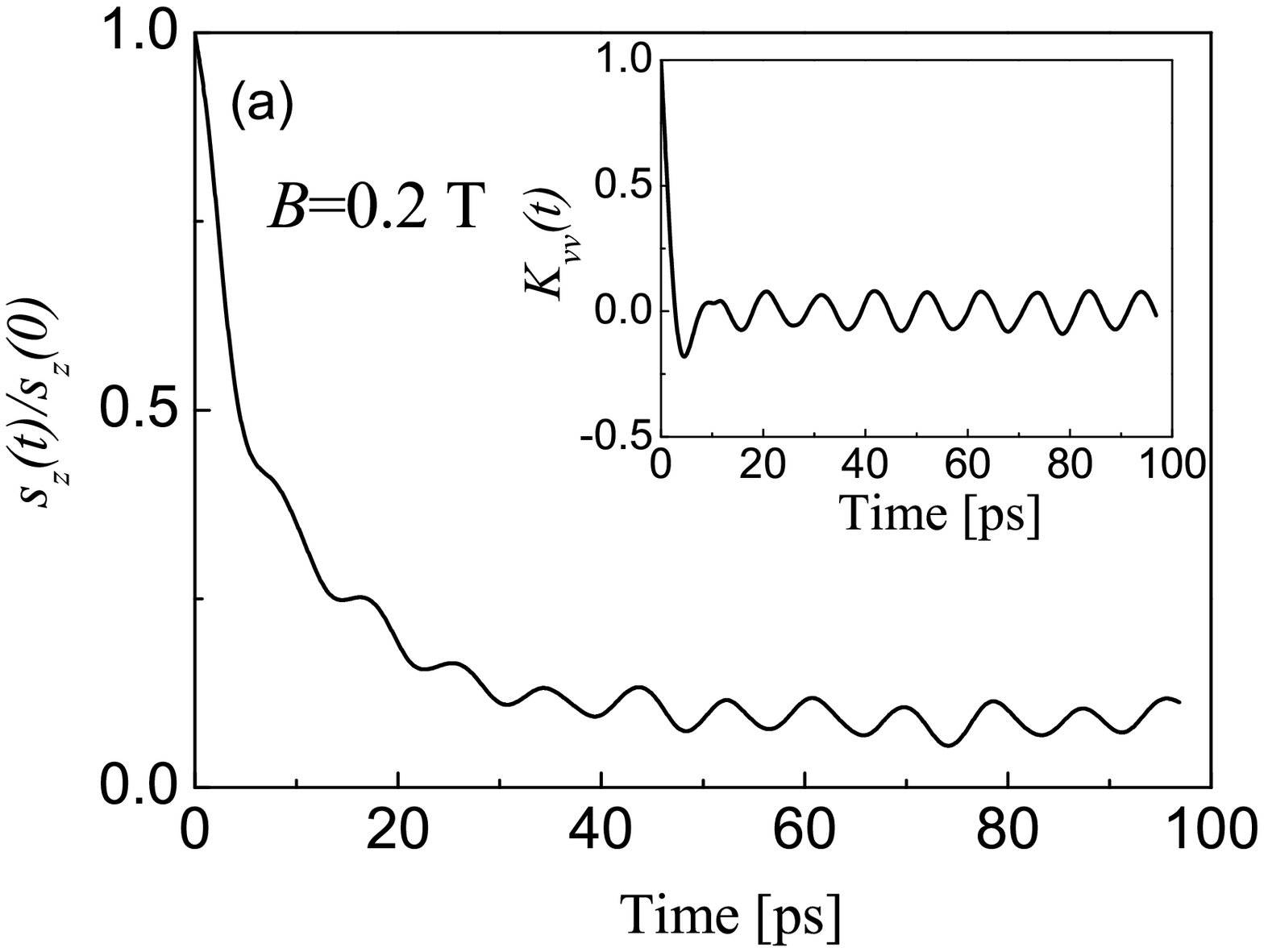}
\includegraphics[width=0.4\linewidth]{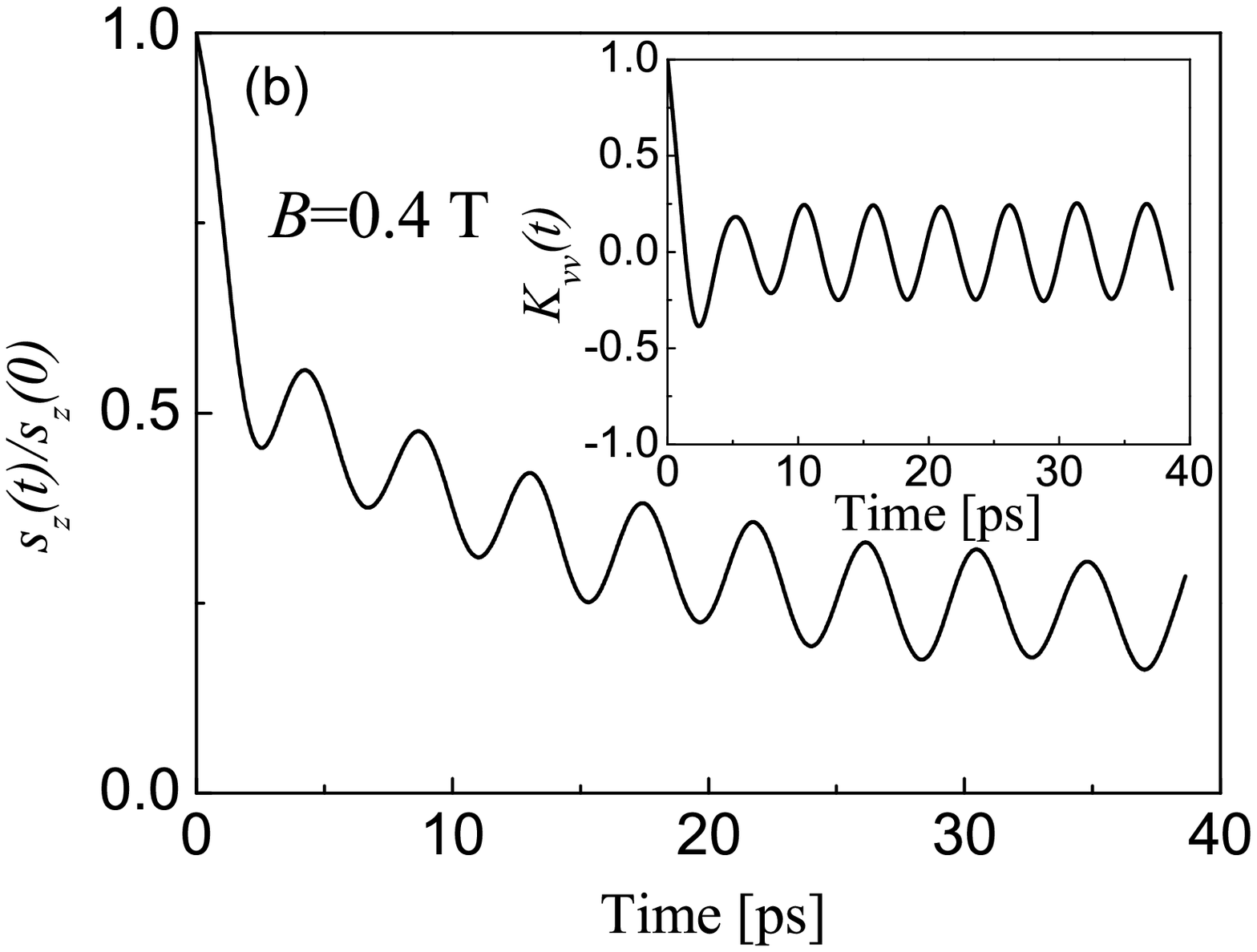}
\caption{The results of the Monte-Carlo simulation for 4096
electrons of the spin dynamics $s_z(t)/s_z(0)$ and the velocity
autocorrelation function $K_{vv}(t)$. The scatterer radius is
$a=2\times 10^{-6}$~cm, $v=2\times 10^{7}$ cm/s, $N=3\times
10^{9}$ cm$^{-2}$,  and $B=0.2$~T, $\Omega_0=4\times 10^{11}$
s$^{-1}$ [panel (a)] and $B=0.4$~T, $\Omega_0=8\times 10^{11}$
s$^{-1}$ [panel (b)]; $\omega_c=3.0\times 10^{12}$ s$^{-1}$/T.}
\label{fig:montecarlo}
\end{figure*}

Therefore, if in the 2DEG the transitions between circling and
wandering trajectories are strictly forbidden, at $t\gg \tau
_{s}(B)$ (defined by Eq. \eqref{taus}) the total spin does not
decay to zero as it was assumed previously, but oscillates around
the non-zero value \
\begin{equation}
s_{z}^{t}=s_{0}P_c (B)\frac{\omega _{c}^{2}}{\omega _{c}^{2}+\Omega _{0}^{2}}%
,  \label{sztail}
\end{equation}%
where $s_{0}$ is the initially excited spin. Further we consider
the case of the strong enough magnetic fields, $\omega _{c}\gg
\Omega _{0}$ where the oscillations of the circling electron spin
are small and the spin relaxation is followed by the non-vanishing
tail with the magnitude $s_{z}^{t}=P_c (B)s_{0}$. In order to have
an insight into the tail formation we performed a Monte-Carlo
simulation of the electron transport and spin dynamics with the
results shown in Fig.~\ref{fig:montecarlo}. It is seen that the
exponential decay of the velocity autocorrelation function
$K_{vv}(t)=\langle{\bf v}(t){\bf v}(0)\rangle/v^2$ (shown in the
inset) is followed by the oscillations due to the circling
electrons. The magnitude of the oscillations corresponds to the
value of $P_c (B)$. The
polarization exhibits oscillatory tail in agreement with Eqs. %
\eqref{szt}, \eqref{sztail}. We mention that in nonquantizing
field $B=0.5$~T the tail has a considerable magnitude $P_c(B)>0.1$
already at low mobilities $\mu =5\times 10^{4}$ cm$^{2}/$Vs.

Now we turn to the decay of the spin polarization tail. Any
additional scattering process such as inelastic scattering or
scattering by a random potential destroys the tail. With the
relaxation time approximation the $z$ spin component is
described by the coupled kinetic equations:%
\begin{eqnarray}
\frac{\partial s_{z}^{c}}{\partial t} &=-&\frac{s_{z}^{c}[1-P_c
(B)]-s_{z}^{w}P_c (B)}{\tau _{j}(B)},  \label{system_c} \\
\frac{\partial s_{z}^{w}}{\partial t} &=&\frac{s_{z}^{c}[1-P_c
(B)]-s_{z}^{w}P_c (B)}{\tau _{j}(B)}-\frac{s_{z}^{w}}{\tau
_{s}(B)}.  \notag \label{system_w}
\end{eqnarray}%
Here superscript $c$ ($w$) corresponds to circling (wandering)
electrons, respectively, and $\tau _{s}(B)$ and $\tau _{j}(B)$ is
the spin relaxation time for wandering electrons and the
intermixing time between the closed and wandering trajectories.
The first term in the numerator of the right hand side of the Eqs.
\eqref{system_c} denotes the transfer of the electron from the
circling to the wandering trajectory and is proportional to the
scattering intermixing rate and the fraction of the wandering
trajectories $1-P_c(B)$ and the second term
refers to the inverse process. In deriving Eqs. %
\eqref{system_c} we have neglected small oscillations of $%
s_{z}^{c}$ and took into account D'yakonov-Perel' spin relaxation
of the wandering electrons (second term of the second line in
\eqref{system_c}).
 The eigenrates of the system $\tau^{-1}_{1,2}(B)$ are
\begin{equation}
\frac{1}{\tau _{1,2}(B)}=\frac{1}{2}\left( \Gamma _{js\text{ }}\pm \sqrt{%
\Gamma _{js\text{ }}^{2}+4\frac{P_c (B)-1}{\tau _{j}(B)\tau
_{s}(B)}}\right) ,  \label{rates}
\end{equation}%
where $\Gamma _{js\text{ }}=\tau^{-1}_{j}(B)+\tau^{-1}_{s}(B).$
Thus, the total spin of electrons $s_z(t)$ relaxes according to
the two exponential law: $s_z(t) =
\alpha\exp{(-t/\tau_1(B))}+\beta\exp{(-t/\tau_2(B))}$ where
coefficients $\alpha$ and $\beta$ depend on the initial condition
and $B$. Since $\tau_2(B)\geq \tau_1(B)$ long-time $s_z(t)$
asymptotics is determined by $\tau_2(B)$ which can be identified
with the tail relaxation time.

For example, in the limit of weak magnetic fields $P_c (B)\ll 1$
and if $\tau _{s}(B)\ll \tau _{j}(B)$ the relaxation times are
$\tau _{1}(B)=\tau _{s}(B)$ and $\tau _{2}(B)=\tau _{j}(B)$ and
the total electron spin rapidly decays to the tail value $P_c
(B)s_{0}$ and then slowly relaxes during the time $\tau _{j}(B)$
(as almost every "additional" scattering transfers electron from
the circling to wandering trajectory where the spin is quickly
lost).

\begin{figure*}[htbp]
\includegraphics[width=0.4\linewidth]{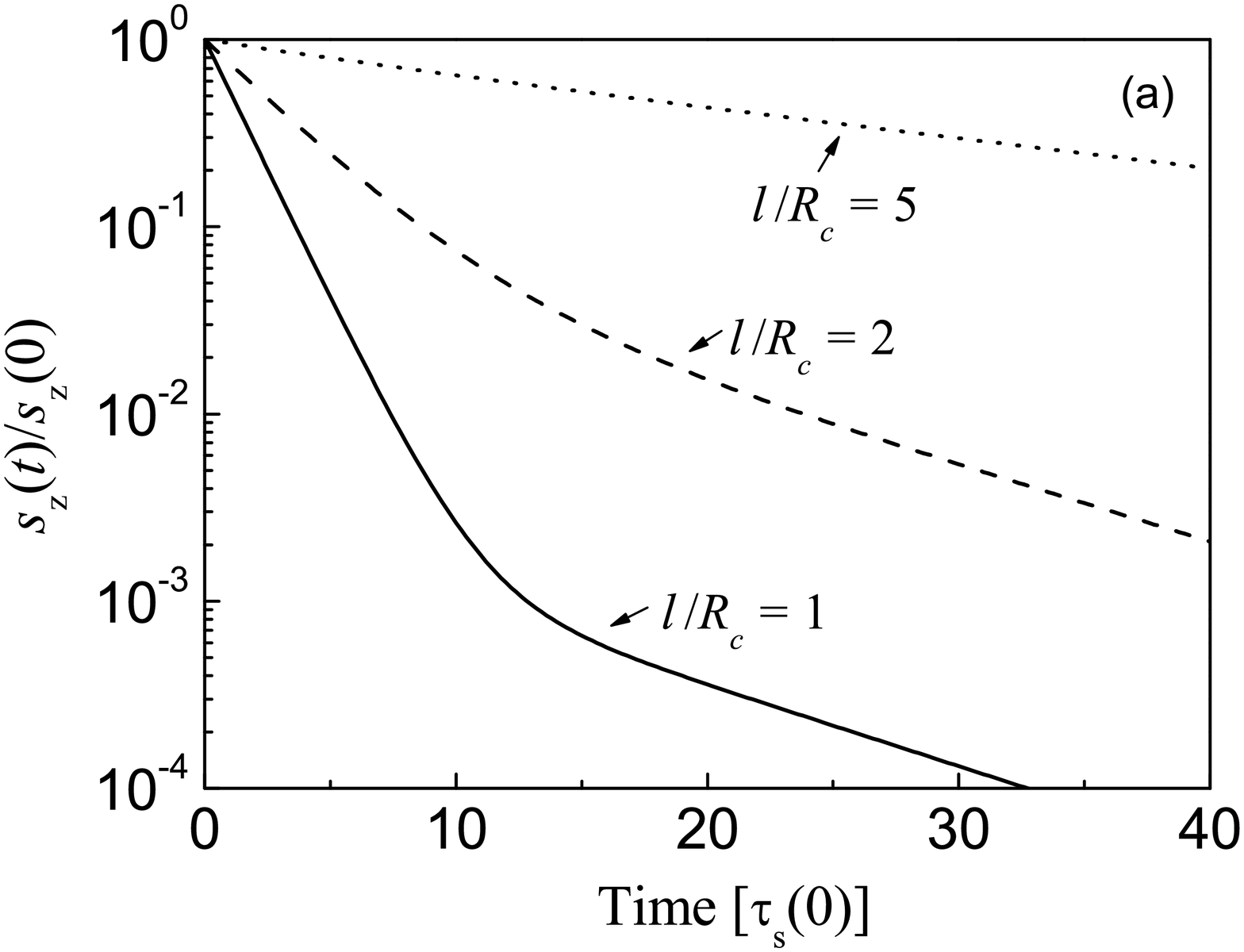}
\includegraphics[width=0.4\linewidth]{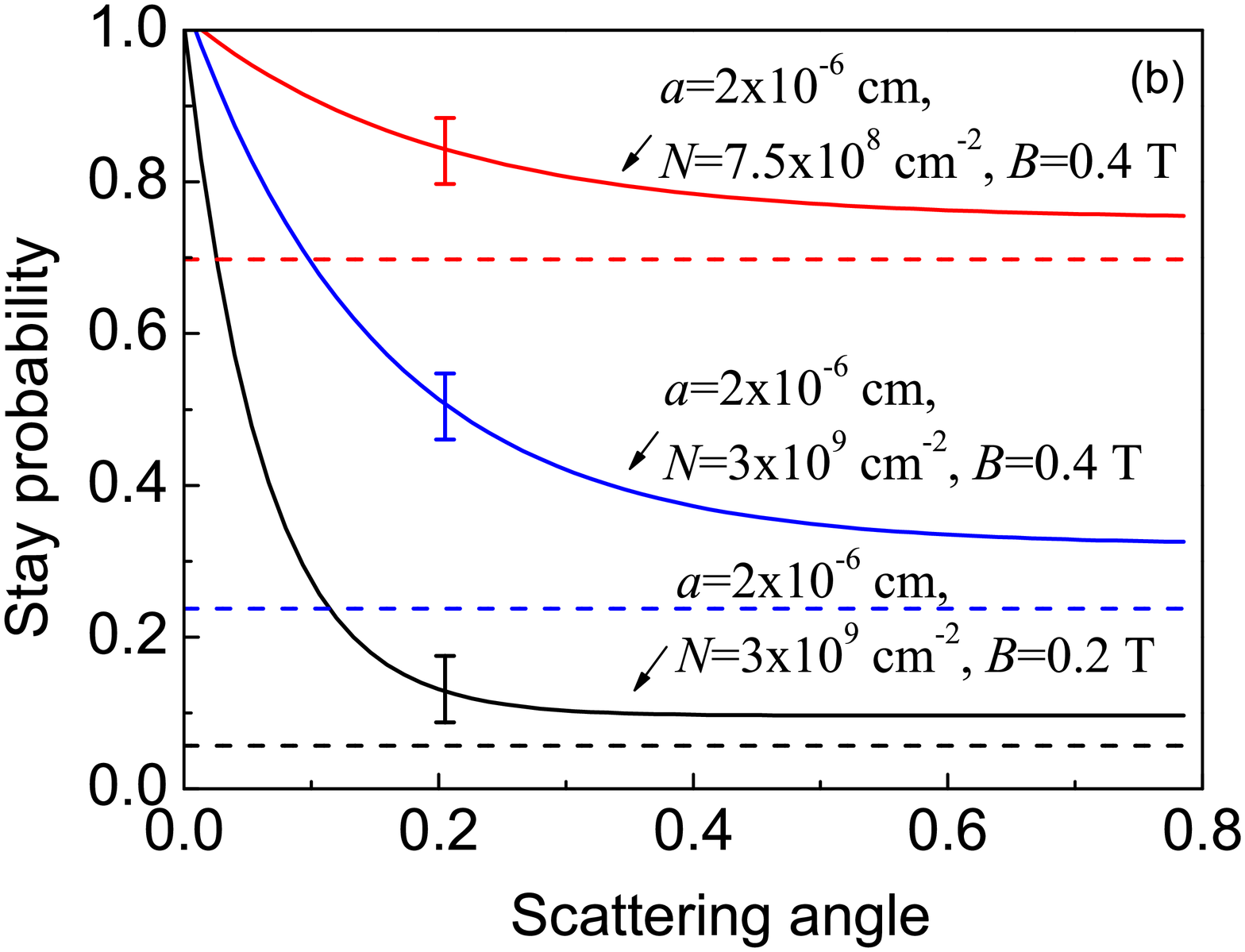}
\caption{{(a) Electron spin $s_z$ time dependence calculated from
Eqs. \eqref{system_c}
for the different values of the magnetic field: $%
l/R_c = 1,2,5$ from the bottom to the top, $\tau_j =
10\tau_s(B=0)$. (b) The scattering probability found by the
Monte-Carlo simulation for (solid lines) and values of $P_c(B)$
(dashed). The parameters used in the calculations are shown above
each curve. Other parameters are same as in Fig.(2). The accuracy
of the Monte-Carlo computation is shown by the error bars.}
}\label{fig:inelast}
\end{figure*}

When the magnetic field becomes stronger with $\omega _{c}\tau \gg
1,$ $\tau _{s}(B)=\tau _{s}(0)\left( \omega _{c}\tau \right) ^{2}$
increases quadratically with the field and the fraction $P_c
(B)=1-3\pi /2\omega _{c}\tau $  tends to unity. Two limiting
regimes as a function of $B$ are possible. In weak fields, where
the intermixing $\tau_j$ is still much longer than $\tau _{s}(B)$,
the spin tail relaxation time is field independent. With the
increase of $B$, the intermixing time becomes smaller than $\tau
_{s}(B)$, and, therefore, the tail disappears. In such a case only
a small fraction $\left( \sim \omega _{c}\tau \right) $ of
electrons is loosing at a given moment their spin, however
circling and wandering electrons are intermixed on a short
timescale.
Therefore the tail is not observed but the spin relaxation time is:%
\begin{equation}
\tau _{2}(B)\approx \frac{l}{2\pi R_{c}}\tau _{s}(B)\approx
\frac{2(\omega _{c}\tau )^{3}}{3\pi \Omega _{0}^{2}\tau} ,
\label{tail_strong_wp1}
\end{equation}%
being proportional to $B^{3}$ (contrary to the standard result $B^{2}$~\cite%
{ivchenko73}). Here $\tau _{2}(B)$ does not depend on the
inelastic scattering time despite the presence of the strong
inelastic scattering which is necessary here only to intermix the
trajectories sufficiently quickly. These results are shown in
Fig.~\ref{fig:inelast}a  illustrating the evolution of the spin
relaxation process $s_z(t)$ and the rate with the change in the
number of wandering trajectories governed by the $B$-dependent
ratio $l/R_c$.

In sufficiently strong fields, the number of trajectories
colliding several times with the same impurity, increases, and at
a critical field such as $4\pi N(a+R_c)^{2} \approx 4.4$ and
$(R_c/l)^{2}=4.4Na^2/\pi\ll 1$ \cite{baskin98} all the paths
become confined. For the parameters considered in Fig.(2), this
corresponds to $B>0.8$ T. In this case the spin relaxation is
strongly suppressed \cite{lyubinskiy06}. The detailed analysis of
this regime is beyond the scope of this Letter.

Below, we consider two kinds of the microscopic processes which
contribute to the intermixing time $\tau_j(B)$ between closed and
wandering trajectories. Namely, we discuss the destruction of the
tail due to: (i) presence of the smooth potential, (ii) inelastic
processes.

(i) In the real QWs the typical disorder can be considered as the
superposition of the short-range scattering centers
(\footnote{These centers can be introduced artificially as
antidotes, see e.g. \cite{pershin04,polyakov:205306,gusev02}}) and
the long-range smooth potential of the remote dopants and
interface roughness~(\footnote{The monolayer fluctuations of the
QW interfaces induce the potential $\delta V\sim 1\ldots 10$~meV
with the lateral size $50\ldots 100$~\AA. Thus, the scattering
length on the interface roughnesses $a$ is of the order of
$10^4$~\AA}). The correlation radius of this potential $d$ is the
distance between the dopant layer and the
QW.~\cite{polyakov:205306,mirlin:126805} We will be interested in
the weak smooth disorder where the transport scattering time by
long-range potential  $ \tau _{l}$ satisfies the condition $\tau
_{l}\omega _{c}\gg 1,$ and, therefore, the circling trajectories
remain at least approximately intact. The displacement of the
orbit center in the smooth potential provided $R_c\gg d$ is
$\delta r \sim l_{l}/(\omega _{c}\tau _{l})^{3/2} \sim
\sqrt{R_c^3/d}(W/E_F)$, where $l_{l}=v_{F}$ $\tau _{l}$ is the
transport mean free path~there, $W$ is the potential amplitude.
\cite{mirlin:2801} The transfer between trajectories is governed
by the ratio $\delta r/a$. The rich transport
\cite{polyakov:205306} and spin dynamics of the electrons in the
system with two types of disorder are beyond the scope of this
Letter. Here we consider the simplest situation where $\delta r
\ll a$ and thus $\delta r \ll d$. Therefore, electron drifts in
the potential gradient averaged over its path and sweeps during
$n$ revolutions an area $S\sim 2\pi R_c n \delta r $. It will be
transferred from the circling trajectory to the wandering as soon
as it meets a short-range center in this area, i.e. $SN\sim 1$.
So, the intermixing time can be estimated as
\begin{equation}\label{jump_smooth}
\tau _{j}^{sp}(B) \sim \frac{T_c}{N R_c \delta r } {\propto
B^{3/2}}, \quad \delta r/a \ll 1,
\end{equation}%
where $T_{c}=2\pi /\omega _{c}$ is the cyclotron period and it is
assumed that $a l_{tr}/(dR_c) \ll 1$. For the purpose of the
present paper we note that the electron drift in the smooth
potential leads to the spin relaxation even for the circular
trajectories, therefore the applicability of the Eq.
\eqref{jump_smooth} together with the system \eqref{system_c} is
guaranteed provided if the intermixing time is much longer than
the spin relaxation time of the circling electrons.

 (ii) The inelastic processes (electron-electron or electron-phonon
scattering) change randomly electron wave\-vector and energy and
thus transfer electrons between circling and wandering
trajectories. In the low temperature regime where the Fermi
surface is well defined, $E_F\gg k_B T$, where $E_F$ is the Fermi
energy and $k_B T$ is the temperature measured in the energy units
all the scattering processes are statistically
quasi-elastic~\cite{gantmakher87}, i.e. the transferred energy $
\sim k_B T$ is much smaller than electron kinetic energy $E_F$.
The crucial question is to determine the probability that the
scattering at the angle $\vartheta $ (Fig.~\ref{fig:scatter}) is
not accompanied by the electron transition from the circling to
wandering orbit. From Fig.~\ref{fig:scatter} it is clear that the
displacement of the circular orbit by the distance of the order of
the impurity separation $b$ will lead to the electron collision
with some scattering center thus the probability should have a
peak in the vicinity of $\vartheta =0$ with the angular width of
the order of $b/R_{c}\sim 1/(N^{1/2}R_c)$. At larger angles the
probability saturates at the value close to $P_c (B)$ representing
the fraction of the circling electrons. This prediction is
corroborated by the Monte-Carlo calculation, see
Fig.~\ref{fig:inelast}b.

In a degenerate 2D electron gas with carrier concentration ranging in $%
10^{11}\ldots 10^{12}$~cm$^{-2}$ the Coulomb interaction is
screened and the electron-electron scattering is almost isotropic
and its temperature dependency is determined by the
Pauli-exclusion principle only. At not too low temperatures
$k_{B}T\gtrsim \hbar c_s\sqrt{\langle k_{z}^{2}\rangle }$ where
$c_s$ is the speed of sound, and $\sqrt{\langle k_{z}^{2}\rangle
}$ is the root mean square of the electron wavevector in the
growth direction, the electron-acoustic phonon scattering is
dominated by deformation-potential coupling and is isotropic as
well. Thus we further assume that the inelastic scattering is
isotropic and can be described in the framework of the relaxation
time $\tau _{j}$ which can be estimated
as for electron-electron and electron-phonon scattering, respectively~\cite%
{price81,glazov04a}:
\begin{equation}
\tau _{j}^{ee}=\zeta_{e}\frac{\hbar }{E_{F}}
\left(\frac{E_{F}}{k_{B}T}\right)^{2},
\quad \tau _{j}^{ph}=\zeta_{ph}\frac{\rho c_s^{2}\hbar ^{3}}{D^{2}mk_{B}T%
\sqrt{\langle k_{z}^{2}\rangle }}.  \label{tau_j}
\end{equation}%
Here $\zeta_{e}=4/[\pi \ln (E_{F}/k_{B}T)]$
for Fermi level electrons, $D$ is the deformation potential constant, $%
\rho $ is the material density, and $\zeta_{ph}\sim 1$ is the
numerical coefficient. At lower temperatures the electron-phonon
scattering is partially suppressed due to Pauli-exclusion
principle and by the phonon-freeze-out and the coefficient
$\zeta_{ph}$ becomes strongly temperature dependent. Its exact
calculation represents an extremely difficult numerical tasks but
the simple estimates show that in the temperature range $T \leq
20$~K and for electron concentration $N_e\sim 10^{11}$ cm$^{-2}$
the electron-electron scattering is dominant. We note that the
fraction of the circling electrons $P_c(B)$ Eq. \eqref{wp} has a
purely geometric character, thus the tail is insensitive to the
thermal broadening of the electron distribution provided inelastic
processes are weak enough.

Finally, we would like to comment on the quantum effects neglected
above. First, the quantization of electron orbits leads to the
oscillatory $B$-dependence of the spin relaxation rate
\eqref{taus} ~\cite{burkov:245312}. This effect is small for high
Landau levels ($E_{F}/\hbar \omega _{c}\gg 1$) and vanishes as
temperature increases. Second, the quantization of the electron
orbits modifies the relaxation rates of inelastic scattering in
Eq. \eqref{tau_j}. However, for semiclassical Landau levels this
modification is negligible~\cite{gantmakher87}. Third, the
magnetic field leads to the spin polarization either due to Zeeman
effect or as a result of the SO coupling and Landau
quantization~\cite{rashba64}. The former contribution is
proportional to $g\mu_{B} B/E_F$ where $g$ is the electron
$g$-factor and $\mu_B$ is Bohr magnetron. For the typical GaAs
structures $g$  is small, besides it can be adjusted to zero with
the structure parameters~\cite{ivchenko05a}. The contribution
$|\mathcal{P}_{L}|=|\langle\psi_{L,\pm}|\sigma_z|\psi_{L,\pm}\rangle|$
of the latter effect can be evaluated for $L$th Landau level
$E_{L,\pm}=\hbar\omega_{c}(L\pm\sqrt{\gamma^{2}L+1/4})$ with
$\gamma =\hbar \Omega _{0}/(2\sqrt{E_{F}\hbar \omega _{c}})$ as
$|\mathcal P_L|=\bigl(2\sqrt{\gamma^{2}L+1/4}\bigr)^{-1}$ and is
negligible for high Landau leves with $\gamma^{2}L \gg 1$.

In conclusion, we have theoretically analyzed the spin dynamics of
the 2D electron gas scattered by short-range defects in the
classically strong magnetic fields. We have shown that the tail in
the spin polarization appears as a result of the collisionless
circular motion of the fraction of electrons. It was demonstrated
that the tail decays due to the presence of the additional weak
long-range potential or due to inelastic scattering. In strong
magnetic fields the long-time spin relaxation is slowed down as a
cube of the field. The predicted phenomena open new perspectives
in the spin dynamics control in the 2DEG.

\acknowledgments

The authors are grateful to E.L. Ivchenko and J.E. Sipe for the
valuable discussions. M.M.G. acknowledges the financial support
from RFBR, programs of RAS and ``Dynasty'' foundation -- ICFPM.
E.Y.S. was supported by DARPA SpinS program and the FWF (Austria)
grant P15520.

\end{document}